\documentclass[%
 reprint,
showpacs,preprintnumbers,
 amsmath,amssymb,
 aps,
pra,
]{revtex4-1}

\usepackage{graphicx}% Include figure files
\usepackage{dcolumn}% Align table columns on decimal point
\usepackage{bm}% bold math
\begin{document}

%\preprint{APS/123-QED}

\title{Light-controlled perfect absorption of light.}% Force line breaks with \\

\author{K. Nireekshan Reddy and S. Dutta Gupta}
\email{sdghyderabad@gmail.com}
\affiliation{School of Physics, University of Hyderabad, Hyderabad-500046, India}


%\affiliation{
%\affiliation{
%\affiliation{%


\date{\today}% It is always \today, today,

\begin{abstract}
We study coherent perfect absorption (CPA) of light in a Kerr nonlinear metal-dielectric composite medium, illuminated from the opposite ends. Elementary symmetry considerations reveal that
equality of the incident light intensities is a prerequisite to ensure CPA in both linear and nonlinear
systems for specific system parameters. We also derive the sufficient conditions for having CPA. We further show that while CPA in a linear system is insensitive to the incident
power level, that in a nonlinear system can be achieved only for discrete intensities with interesting
hysteretic response. Our unified formulation of  CPA and waveguiding identifies them as opposite scattering phenomena. We further investigate light-induced CPA in on- and off- resonant systems.
%\begin{description}
%\end{description}
\end{abstract}

\maketitle

%\tableofcontents
%%%%%%%%%%%%% SECTION I
\section{Introduction}\label{sec:sec1}
In recent years there has been a considerable interest in perfect absorption of single and multiple coherent optical waves. The former is achieved in a critically coupled system \cite{cc1,cc2,cc3,cc4}, while the the latter is referred to as coherent perfect absorption (CPA) \cite{cpa1,cpa2,cpa3,cpa4}. Both the phenomena are based on `perfect' destructive interference, demanding equality of the amplitudes of the interfering waves, having a phase difference of odd multiples of $\pi$. Most of the investigations till date focus on linear structures with very few exceptions \cite{n1,n2,n3,n4}. Very recently a Kerr nonlinear system was investigated for power dependence of crtical coupling \cite{ncc}. In this paper we address a more fundamental issue of coherent perfect absorption in a Kerr nonlinear slab. In order to make connection with earlier work on a linear system \cite{cpa2}, we consider the slab of a metal-dielectric composite medium assuming  it to have an overall Kerr nonlinearity. Recall that a rigorous theory for heterogeneous medium with nonlinear inclusions or host exists \cite{tmat,sipe}, which leads to a complex susceptibility $\chi^{(3)}$ having nonlinear absorption as well. In order to retain the simplicity we ignore all such effects and assume the nonlinearity parameter to be a constant, while retaining the absorption/dispersion effects in the linear susceptibility. Note that the theory developed is suitable for any Kerr nonlinear medium with finite linear absorption. We adopt the nonlinear characteristic matrix theory (NCMT) \cite{ncmt}, for example, for s-polarized waves to find the relation between the amplitudes inside and outside of the nonlinear medium. Elementary symmetry arguments for CPA is shown to lead to identical incident intensities from both the ends, enabling one to define the symmetric and antisymmetric CPA conditions. A somewhat different boundary conditions, namely, of null input waves with finite scattered waves lead to the conditions for wave guiding. Obviously for a guide one has to assume the scattering waves to be of evanescnent character. Thus a unified formulation of both CPA and wave guiding reveals yet another striking feature of perfect absorption. Recall that till now there are parallels of CPA referred to as anti-lasing or time reversed lasing \cite{cpa1,al}.
%%%%%%%%%% FIGURE 1
\begin{figure}[ht]
\center{\includegraphics[width=4cm]{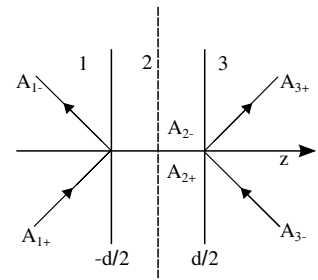}}
\caption{Schematics of the nonlinear system and illumination.} 
\label{fig:fig1}
\end{figure}
%%%%%%%%%%%%%%%
\par
A comparison of the linear and the nonlinear systems is shown to lead to another interesting feature. For example, CPA at a given wavelength in a linear system is insensitive to the intensity, provided that inputs from both the ends match. In a nonlinear system, in contrast, CPA is realizable only at distinct intensity levels. In fact, we derive the sufficient conditions for having CPA. We demonstrate recurrence of the CPA minima in a nonlinear system with an example, where the linear counterpart is tuned at a CPA resonance. Moreover, the system is shown to have hysteretic response, which could be very useful for optical switching, memory and logic operations. We also study the case where the linear counterpart is away from the CPA resonance, we show that for specific detuning, the system can be brought back to CPA by nonlinearity induced changes.
\par 
The organization of the paper is as follows. In Section~\ref{sec:sec2} we formulate the problem and present our analytical results. In Section~\ref{sec:sec3} we present the numerical results and discuss them. We summarize the main results of the paper in Conclusions.
%%%%%%%%%%%%%%%% SECTION II
\section{Formulation of the problem}\label{sec:sec2}
We consider the system shown in Fig.~\ref{fig:fig1}, where a Kerr nonlinear slab of width $d$ is illuminated  at an angle $\theta$ from both the sides by s-polarized plane monochromatic waves (with wavelength $\lambda$). The dielectric function for the nonlinear slab is assumed to have the following form
\begin{eqnarray}\label{eq1}
\epsilon_2=\bar{\epsilon}+\alpha |E|^2,
\end{eqnarray}
where the background material is a metal dielectric composite with dielectric constant $\bar{\epsilon}$ and $\alpha$ is the nonlinearity constant. Using Bruggeman formulation the dielectric constant $\bar{\epsilon}$ can be expressed as \cite{cpa2}
\begin{widetext}
\begin{equation}\label{eq2}
\bar{\epsilon}= \frac{1}{4}\left\lbrace\left(3f_m -1\right)\epsilon_m +\left(3f_d-1\right)\epsilon _ d
 \pm \sqrt{\left[\left(3f_m-1\right)\epsilon_ m + \left(3f_d -1\right)\epsilon_ d\right]^2 + 8\epsilon_m\epsilon_d}\right\rbrace,
\end{equation}
\end{widetext}
where, $f_m$ and $\epsilon_m$ ( $f_d$ and $\epsilon_d$ ) are the volume fraction and the permittivity of the metal (dielectric), respectively, with $f_d=1-f_m$. In order to ensure causality, the square root is computed such that imaginary part of $\bar{\epsilon}$ is always positive. In  Eq.~(\ref{eq1}) dispersion is included in the linear part of $\bar{\epsilon}$, while it is neglected in the nonlinearity constant $\alpha$. As mentioned in the Introduction, a more rigorous approach requires the inclusion of dispersion in both, as discussed by Agarwal \textit{et al.} \cite{tmat}. The general solutions for such a nonlinear slab are not known even for s-polarized light. Recall that there exists exact theory for a Kerr nonlinear slab which is also quite involved \cite{chen}. In contrast, for weakly nonlinear systems,  a simple NCMT can be used to capture all the related phenomena.
Recall that NCMT is based on the slowly varying envelope approximation and has proved to be very effective in easy probing of the nonlinear effects in a variety of situations like weak photon localization \cite{loc}, gap solitons in periodic structures \cite{gaps}, optical  bistability with Fabry-Perot or the surface modes \cite{ncmt,fp1,sm1,sm2}. In the frame work of such a theory the field in the nonlinear medium is expressed as a superposition of forward and backward propagating waves, albeit with power-dependent propagation constants \cite{fp1,mar}
\begin{eqnarray}\label{eq3}
E_{2y}=\tilde{A}_{2+}e^{ip_{2z+}\bar{z}}+\tilde{A}_{2-}e^{ip_{2z-}\bar{z}},
\end{eqnarray}
where $\bar{z}=k_0z$ is the dimesionless length and $p_{2z\pm}$ are the normalized (to $k_0$ ) propagation constants for the forward and the backward waves, given by \cite{fp1,mar}
\begin{eqnarray}\label{eq4}
p_{2z\pm}=\sqrt{\bar{\epsilon}-{p_{x}}^2+|A_{2\pm}|^2+2|A_{2\mp}|^2},
\end{eqnarray}
where ${p_{x}}=\sqrt{\epsilon_0} \sin\theta$, $A_{2\pm}=\sqrt{\alpha}\tilde{A}_{2\pm}$ are the dimensionless amplitudes in the nonlinear medium. The corresponding expressions for the magnetic field can be calculated using Maxwell equations. The next step is to use the boundary conditions for the tangential components of the electric and the magnetic fields at $\bar{z}=\pm k_0d/2$. This yields the following equations.
\begin{eqnarray}
\begin{pmatrix}
A_{1+} \\
A_{1-} \end{pmatrix} &=& {\tilde{M}_I}^{-1} \tilde{M}\left(-\bar{d}/2\right)\begin{pmatrix}A_{2+} \\ A_{2-}\end{pmatrix},\label{eq5}\\
\begin{pmatrix}
A_{3+} \\
A_{3-} \end{pmatrix} &=&{\tilde{M}_I}^{-1}\tilde{M}\left(\bar{d}/2\right)\begin{pmatrix}A_{2+} \\ A_{2-}\end{pmatrix}, \label{eq6}
\end{eqnarray}
where $\bar{d}=k_0d$ is the dimensionless width of the slab and  $\tilde{M_I}$ and  $\tilde{M}\left(\bar{d}/2\right)$ are given by 
\begin{widetext}
\begin{equation}
\tilde{M_I}={\begin{pmatrix}
1 & 1 \\
p_{1z} & -p_{1z} \end{pmatrix}},\hspace{0.1cm} \tilde{M}\left(\bar{d}/2\right)=\begin{pmatrix} \exp{(ip_{2z+}\bar{d}/2)} & \exp{(-ip_{2z-}\bar{d}/2)} \\ p_{2z+}\exp{( ip_{2z+}\bar{d}/2)} & -p_{2z-}\exp{(-ip_{2z-}\bar{d}/2)} \end{pmatrix},\label{eq7}  
\end{equation}
\end{widetext}
 Using Eqs.~(\ref{eq5})-(\ref{eq7}) it can be easily shown that in order to have CPA ($A_{1-}=0, A_{3+}=0$) the incident intensities must be the same. One can then talk about the symmetric and the antisymmetric field distributions in the nonlinear system. For example, for the symmetric (antisymmetric) mode $A_{2-}=A_{2+}=A$ ($A_{2-}=-A_{2+}=-A$) with the consequent implications that $A_{3-}=A_{1+}=A_{in}\hspace{0.05cm}$($A_{3-}=-A_{1+}=-A_{in}$). One can then reduce Eqs. (\ref{eq5})-(\ref{eq7}) to the following form
\begin{eqnarray}\label{eq8}
\begin{pmatrix}
1 & \pm1\\ p_{2z} & \mp p_{2z}
\end{pmatrix} \begin{pmatrix}
A \\ A
\end{pmatrix}&=&M_{d/2}\begin{pmatrix}
1 \\ -p_{1z}
\end{pmatrix}A_{in},
\end{eqnarray} 
 where $p_{2z+}=p_{2z-}=p_{2z}$, and $M_{d/2}$ is the characteristic matrix of the nonlinear half slab \cite{ncmt}. The general form of Eq.~(\ref{eq8}) includes the limiting case of a linear system when $\alpha=0$ and $p_{2z}=\sqrt{\bar{\epsilon}-{p_{x}}^2}$. Note that the case of a linear or Kerr nonlinear waveguide can be obtained using the same procedure with null input and finite evanescent $\left(p_{1z}=i\tilde{p},~ \tilde{p}=\sqrt{{p_{x}}^2-\epsilon_0}\right)$ output, which reduces Eqs.~(\ref{eq5})-(\ref{eq7}) to the following 
\begin{eqnarray}\label{eq9}
\begin{pmatrix}
1 & \pm1\\ p_{2z} & \mp p_{2z}
\end{pmatrix} \begin{pmatrix}
A \\ A
\end{pmatrix}&=&M_{d/2}\begin{pmatrix}
1 \\ p_{1z}
\end{pmatrix}A_{t}.
\end{eqnarray} 
A comparison of Eqs.~(\ref{eq8}) \& (\ref{eq9}) reveal the contrast between CPA and guided wave phenomena as scattering events. The first (second) one corresponds to null output (input) with finite input (output).
Eq.~(\ref{eq8}) can be expressed as 
\begin{eqnarray}
D_S=p_{1z}+ip_{2z}\tan{\left(p_{2z}\bar{d}/2\right)}=0,\label{eq10}\\
D_A=p_{1z}-ip_{2z}\cot{\left(p_{2z}\bar{d}/2\right)}=0,\label{eq11}
\end{eqnarray}
for the symmetric and the antisymmetric modes, respectively. Analogous equations (referred to as the dispersion relations) for the wave guide are given by
\begin{eqnarray}
\tilde{p}_{1z}-p_{2z}\tan{\left(p_{2z}\bar{d}/2\right)}=0,\label{eq12}\\
\tilde{p}_{1z}+p_{2z}\cot{\left(p_{2z}\bar{d}/2\right)}=0.\label{eq13}
\end{eqnarray}
These results tally with the known mode dispersion relations for a symmetric optical waveguide \cite{wg}. Eqs.~(\ref{eq10}) and  (\ref{eq11}) represent the sufficient conditions for the CPA resonance with their roots (if they exist) giving the location and characteristics of the CPA minima. For a linear system a given CPA resonance is insensitive to the input intensities. The situation is drastically different for a nonlinear system, where the scaled propagation constant is intensity dependent. Thus, the relation can be met only at distinct power levels for a given wavelength. Henceforth, we treat the amplitude $A$ as a parameter and trace out the dependence of $|A_t|$ as a function of $|A_{in}|$ for on- and off- resonant situations. 
%%%%%%%%%%% FIGURE 2
\begin{figure}[t]
\center{\includegraphics[width=6cm]{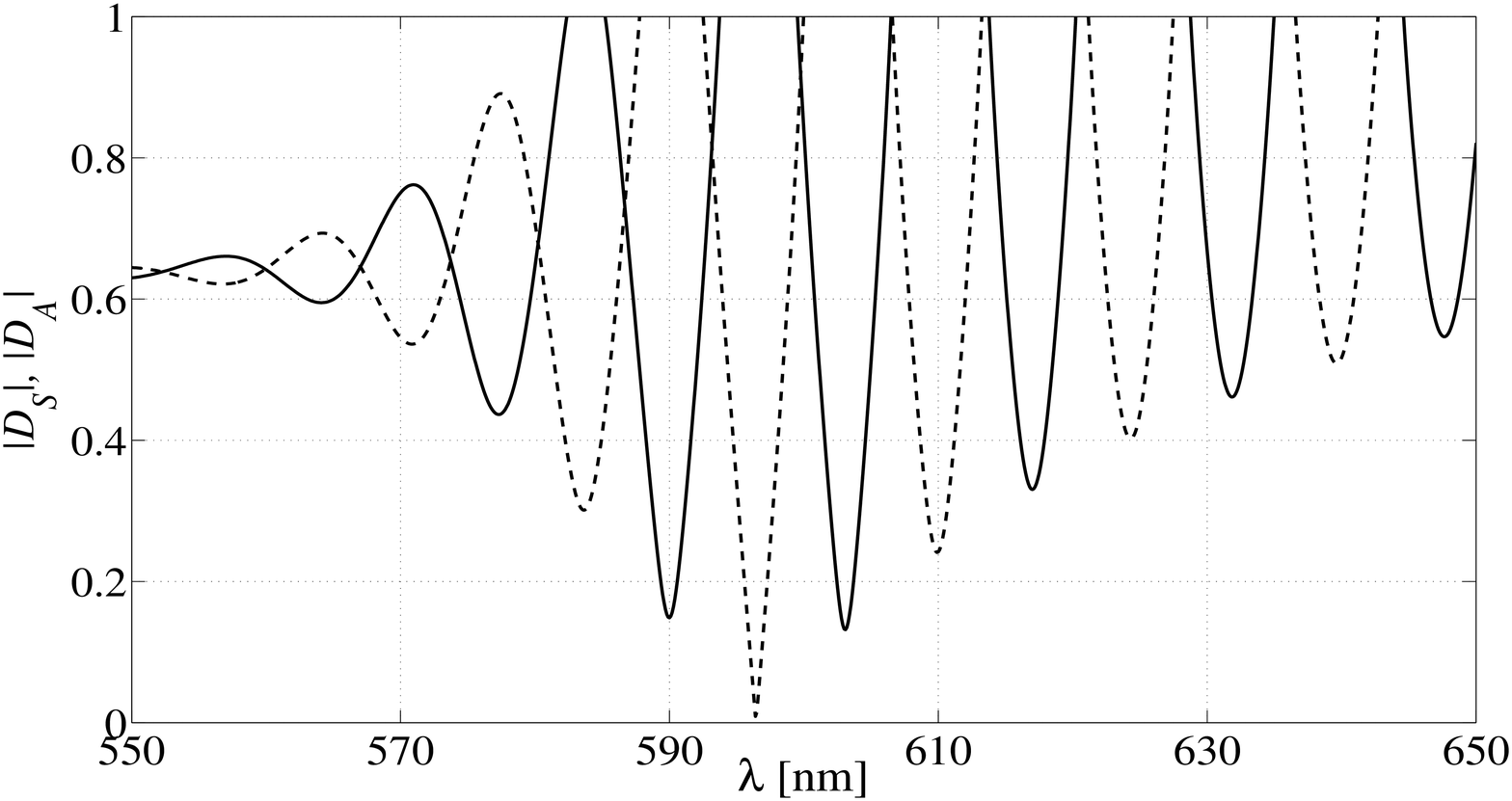}}
\caption{$|D_S|$ (dashed), $|D_A|$ (solid) as functions of $\lambda$, for a linear system with $d=18.58~\mu m$. Other parameters are as follows $\epsilon_d=2.25$, $f_m=0.004$, $\theta=45\,^{\circ}$.} 
\label{fig:fig2}
\end{figure}
%%%%%%%%%%%%%%
\section{Numerical results}\label{sec:sec3}
\begin{figure}[htp]
\center{\includegraphics[width=8.5cm]{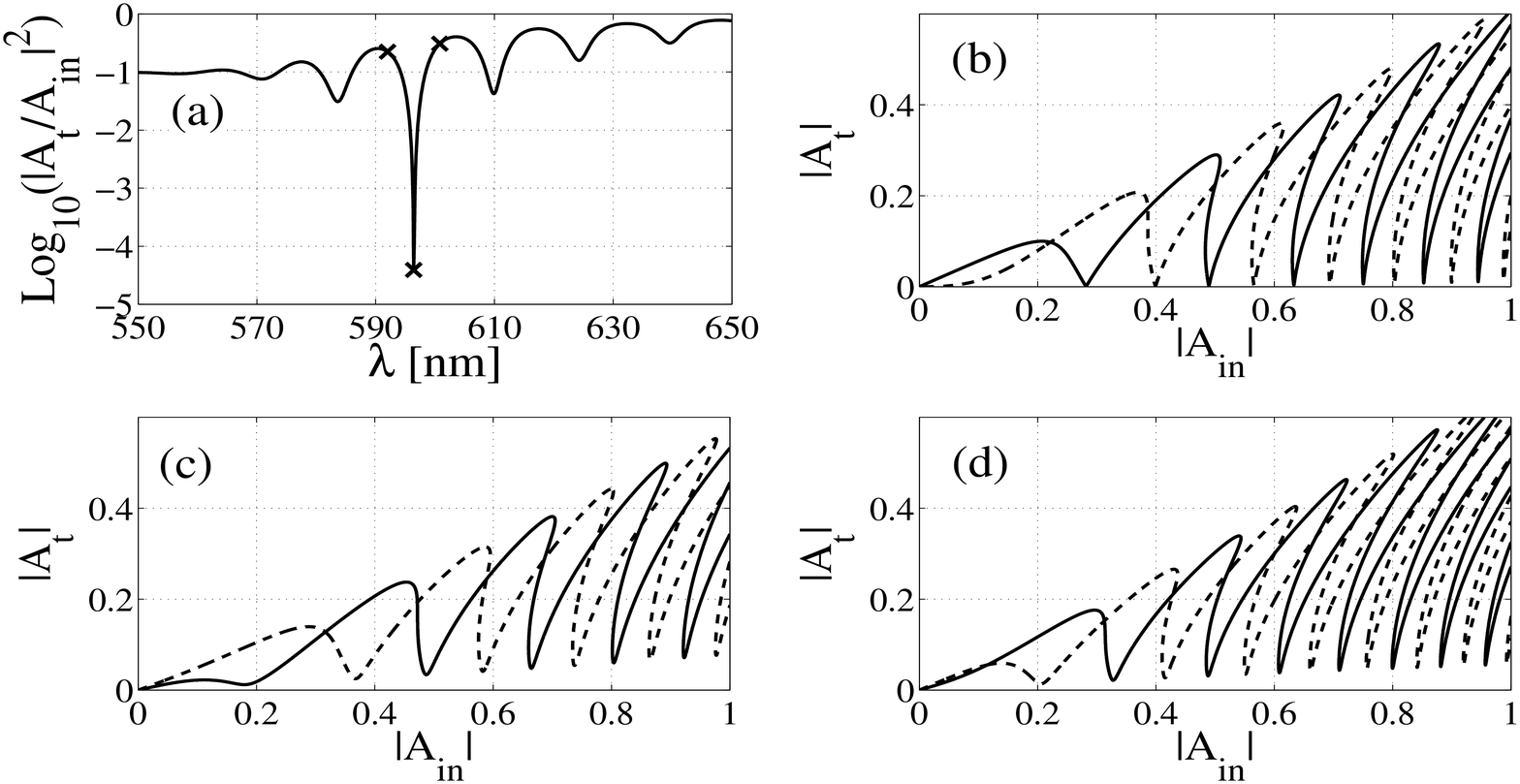}}
\caption{(a) Intensity scattering $\log_{10}|A_t/A_{in}|^2$ as a function of $\lambda$ for $d=18.58~\mu m$. Absolute values of the scattered amplitude $A_t$ as functions of $|A_{in}|$ for (b) $\lambda=596.4nm$, (c) $\lambda=592.0~nm$, (d) $\lambda=600.8~nm$. The dashed (solid) curves are for the symmetric (antisymmetric) modes. Other parameters are same as in Fig.~\ref{fig:fig2}. } 
\label{fig:fig3}
\end{figure}

In this Section we present the numerical results. For most of our calculations, we have taken the following parameters $\epsilon_1=\epsilon_3=1$, $\epsilon_d=2.25$, $f_m=0.004$, $\theta=45\,^{\circ}$, $d=18.5~ \mu m$ and the parameters for the metal (gold) are taken from the experimental work of Johnson and Christie \cite{j&c}. All the media are assumed to be non magnetic. We work away from the localized plasmon resonance and in the range of wavelengths considered Eq.~(\ref{eq11}) can not be satisfied for a linear system (except for the trivial solution). Thus the only possible solution for a linear system corresponds to CPA due to the symmetric mode (Eq.~(\ref{eq10})) at $\lambda=596.4~nm$. This is shown in Fig.~\ref{fig:fig2}, where we have plotted $|D_S|$ (dashed line) and $|D_A|$ (solid line) as functions of $\lambda$. Further for the linear system the absolute value of the normalized scattered amplitude $|A_{t}/A_{in}|$ as a function of wavelength is shown in Fig.~\ref{fig:fig3}(a), which clearly agrees with the minimum in Fig.~\ref{fig:fig2}. 
%%%%%%%%% FIGURE 3   
%%%%%%%%%%%%%%%
The nonlinear responses are shown in Figs.~\ref{fig:fig3}(b)-(d), for wavelengths marked in Fig.~\ref{fig:fig3}(a) by crosses, namely, for $\lambda=596.4~nm$, $\lambda=592.0~nm$ and $\lambda=600.8~nm$, respectively.  It is clear from Fig.~\ref{fig:fig3}(b) that CPA in a nonlinear system is achievable at discrete power levels, if the wavelength corresponds to the CPA minimum of a linear system. In case of wavelengths chosen away from the CPA resonance, nonlinearity can not recover the CPA minima (Figs.~\ref{fig:fig3}(c),~(d)). Note also the bistable response for both the symmetric and the antisymmetric branches. Another pertinent feature is the linear limit of low intensities in Fig.~\ref{fig:fig3}(b), which verifies the possibility of CPA with only the symmetric mode, noted in Fig.~\ref{fig:fig2}.
\par
\begin{figure}[]
\center{\includegraphics[width=8.5cm]{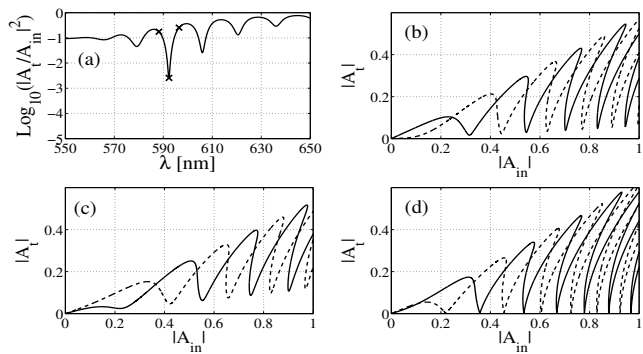}}
\caption{(a) Same as in Fig.~\ref{fig:fig2}(a), except that $d=18.00~\mu m$, (b) $\lambda=592.3~nm$, (c) $\lambda=588.2~nm$, (d) $\lambda=596.4~nm$. } 
\label{fig:fig4}
\end{figure}

We next consider a linear system away from the CPA resonance for $d=18.00~\mu m$, which can not exhibit near total suppression of scattering (see Fig.~\ref{fig:fig4}(a)). We investigate if nonlinearity induced changes can tune the system back to CPA resonance. Indeed, nonlinearity can restore CPA (Fig.~\ref{fig:fig4}(d)) for wavelengths red-detuned from the minimum, while wavelengths corresponding to minimum or a blue-detuning lead to finite scattering Fig.~\ref{fig:fig4}(b),~(c). 
%%%%%%%%%%%% FIGURE 4
%%%%%%%%%%%%%% SECTION IV
\section{Conclusion}\label{sec:sec4}
In conclusion, we have studied a metal-dielectric composite slab with Kerr type nonlinearity under illumination from both sides. We have shown that identical intensities are needed in order to suppress scattering from the system, enabling one to define the symmetric and antisymmetric modes. Intensity dependence of the scattered light as a function of input is shown to lead to CPA at a discrete intensity levels. A unified approach for both CPA and waveguiding leads to yet another fundamental aspects of perfect absorption. Both on- and off- resonant systems are studied for intensity dependence. It is demonstrated that CPA can be restored in an off-resonant system by a suitable control of input light intensity. Our results can find applications in a variety of problems in nanophotonics, plasmonics and solar power harvesting. 
\end{document}